# Making Real Memristive Processing-in-Memory Faster and Reliable

Shahar Kvatinsky
Viterbi Faculty of Electrical and Computer Engineering
Technion – Israel Institute of Technology
Haifa, Israel 3200003
shahar@ee.technion.ac.il

*Abstract*— Memristive technologies are attractive candidates to replace conventional memory technologies, and can also be used to perform logic and arithmetic operations using a technique called 'stateful logic.' Combining data storage and computation in the memory array enables a novel non-von Neumann architecture, where both the operations are performed within a memristive Memory Processing Unit (mMPU). The mMPU relies on adding computing capabilities to the memristive memory cells without changing the basic memory array structure. The use of an mMPU alleviates the primary restriction on performance and energy in a von Neumann machine, which is the data transfer between CPU and memory. Here, the various aspects of mMPU are discussed, including its architecture and implications on the computing system and software, as well as examining the microarchitectural aspects. We show how mMPU can be improved to accelerate different applications and how the poor reliability of memristors can be improved as part of the mMPU operation.

## I. Introduction

Computing systems are typically designed in von Neumann architecture, or an ameliorated version of it, which separates the memory and processing space. In these systems, programs are executed by moving data between the processing unit and memory using specific operations (load/store). While this programming model is simple, the performance of the system is limited by the memory access time, which is substantially higher than the computing time itself. This performance bottleneck (known as the "memory wall") has become even more severe over the years because CPU speed has improved much more than memory speed and bandwidth [1]. Moreover, many modern workloads have high and unstructured data volumes with limited locality, reducing the effectiveness of data caching.

Processing-in-memory (PIM) is an attractive solution to alleviate the memory wall. One such architecture is the *memristive memory processing unit* (mMPU) [2-6]. In the mMPU, memristors are used to construct a dense nonvolatile memory that can also be used to perform *stateful logic* operations, using the same cells. At different times of the program execution, memristors can serve as input, output, latches, and memory cells, enabling real PIM.

In this work, we describe the mMPU architecture, based on a stateful logic [7-8] technique called memristor aided logic (MAGIC) [9-10]. Then, several approaches to improve the performance and reliability of the mMPU are presented.

## II. MAGIC

Several techniques have been proposed to perform logic with memristors [7, 11-15]. *Memristor-Aided loGIC* (MAGIC) [9] is a stateful, in-memory, flexible logic family. In MAGIC, only a single voltage $V_G$ is used to perform a specific function. The initial states (resistance) of the input memristors serve as the input of the logic gate, while the final state (resistance) of the output memristor is the result of the logical operation. Prior to the operation, the output is usually initialized to a known logical state. During the operation, the applied voltage forms a voltage divider and the exact voltage across the output memristor depends on the inputs. Therefore, the output device may switch and by that constitutes the desired logical operation.

The original MAGIC paper [9] presented several functions and concluded that NOR can be performed within the memristive crossbar array. Since then, several additional gates to be performed within the crossbar have been proposed for different memristive technologies [16-17, 21] and were experimentally realized [18, 37]. MAGIC gates can be performed simultaneously on multiple rows/columns within the array. This enables massive parallel execution of different vector operations. The operation of a vector of basic MAGIC NOR gates within a memristive crossbar array is illustrated in Figure 1.

## III. mMPU

The mMPU [2-6] is a standard memristive memory with a few modifications that enable the support of MAGIC-based PIM instructions. In other words, the mMPU functions as a standard memory that supports memory operations (*i.e.*, read and write) with additional PIM capabilities, and thus it is backward compatible with the von Neumann computing scheme. The mMPU architecture is shown in Figure 2. To support PIM instructions, the memory controller [19, 22], the memory protocol [20], and the peripheral circuits (*i.e.*, voltage drivers and row/column decoders) must be modified to support MAGIC instructions [10]. The mapping of data is also modified to maintain persistency and coherence. Note, however, that the memory crossbar array structure itself is not modified and can be in different forms of memristive memory cells, such as 1R (single memristor per cell), 1S1R (1 selector, 1 memristor), and 1T1R (1 transistor, 1 memristor).

## IV. Improving mMPU Performance

The preliminary performance results of the mMPU have shown the potential to improve the throughput and execution time compared to conventional computing systems for different applications. Imani *et al*. demonstrated more than 100X speedup versus GPU for deep neural network training [23]. Haj Ali *et al.* have shown a similar speedup for image processing compared to previously proposed memristive accelerators [2, 24]. The Bitlet model [25] is an analytical model to determine whether a specific application will benefit from the mMPU compared to a von Neumann machine.

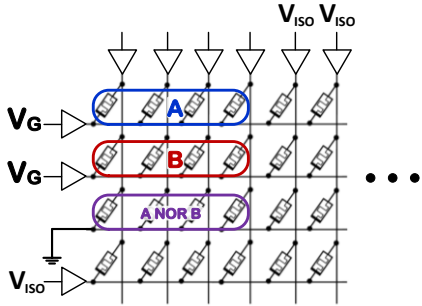

**Figure 1.** A MAGIC NOR operation between two row vectors *A* and *B* is performed within the memristive memory array by applying $V_G$ to the wordlines of the input memristors, ground to the wordline of the output memristor, and $V_{ISO}$ to isolate unselected bitlines and wordlines. The operation takes a single clock cycle regardless of the vector size of *A* and *B*.

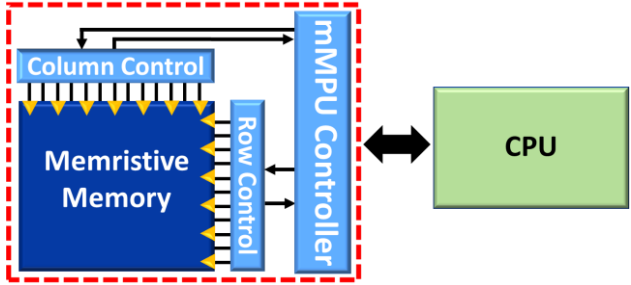

**Figure 2.** Memristive memory processing unit (mMPU). The mMPU is built from memristive memories orchestrated by the mMPU controller. The mMPU supports regular memory operations (read/write), as well as logical operations.

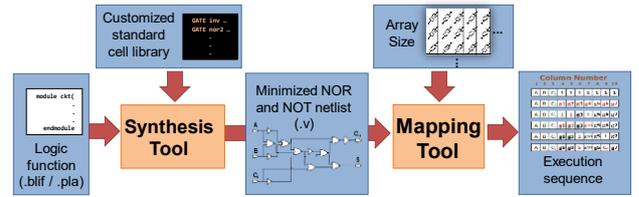

**Figure 3.** SIMPLER MAGIC flow. The desired function is represented as a .blif/.pla file and is generated as a netlist using a standard synthesis tool for a specific technology libraray. Then, the netlist goes into a mapping tool that maps the logic gates into specific memristors in a sequential manner, based on the physical memory array constraints.

The efforts to improve the performance of the mMPU focus on several aspects. The mMPU controller orchestrates the mMPU [19, 26] based on the specific MAGIC operations that are supported. AbstractPIM [27-28] is the first effort to explore the software-hardware interaction of the mMPU, by defining different instruction set architectures (ISA) for different logic primitives and evaluating the performance. For the processor design, several synthesis tools have been proposed to support automatic generation of the mapping and execution sequence for any desired function. SIMPLE [29] is based on solving an optimization to minimize the latency of the execution, while its successor SIMPLER [30] takes a different approach. In SIMPLER, heuristics are used to improve the runtime of the tool. The target of SIMPLER is increasing the throughput, rather than lowering the latency. SIMPLER limits the execution to a single row in the memory array and by that enables execution of multiple operations concurrently. The flow of SIMPLER is shown in Figure 3.

## V. IMPROVING mMPU RELIABILITY

Non-idealities influence the performance and reliability of the mMPU [31-33]. Wald *et al.* [34] explored the influence of process and environmental variation on the execution of MAGIC. Talati *et al.* [35] evaluated the cost of non-ideal interconnect and the cost of internal data movement within the mMPU. Hoffer *et al.* [18] showed experimentally that some memristive technologies fail to execute MAGIC operations due to their voltage threshold values and proposed new gates that are appropriate for the specific memristor properties.

Recently, Leitersdorf *et al.* [36] proposed using internal error correcting codes (ECC) to overcome soft errors in the memristive devices. The difference between standard ECC and the proposed technique is that in the standard manner, the encoding is performed during a read operation, while in the mMPU the aim is to perform the encoding within the memory, without reading the values prior to the logic operations. Leitersdorf *et al.* proposed a diagonal code that can be performed using MAGIC operations as part of the logical execution, and by that improve the reliability of the mMPU and increase the mMPU mean time between failures by nine orders of magnitude.

## VI. CONCLUSION

Processing-in-memory is an attractive approach to overcome the memory wall. However, there are many challenges that need to be considered in order to make PIM systems, such as the mMPU, practical and efficient. These challenges include the entire design stack from the physical non-idealities of the technology up to the software definitions and the programming model.

### ACKNOWLEDGMENT

This research is partially supported by the European Research Council under the European Union's Horizon 2020 Research and Innovation Programme (grant agreement no. 757259) and by the Israel Science Foundation grant no. 1514/17.